\documentstyle[11pt,pazh2,psfig]{article}

\newcommand{\km}{\,\mbox{km}\,\mbox{s}^{-1}}
\newcommand{\erg}{\,\mbox{erg}\,\mbox{s}^{-1}}
\def\Ha{\hbox{H$_\alpha$~}}

\begin{document}

\title{A Giant Bipolar Shell around the WO star in the Galaxy IC 1613: Structure and
Kinematics \thanks{Astronomy    Letters,    Vol.26,No~3, 2000,
pp.153-161.   Translated  from  Pis'ma  v  Astronomicheskii
Zhurnal, Vol.26, No. 3, pp. 190-199. Translated by V. Astakhov.}}

\author{ Afanasiev V.L.\inst{a} \and  Lozinskaya T.A. \inst{b}\thanks{E-mail address for contacts: lozinsk@sai.msu.su}
\and Moiseev A.V. \inst{a} \and  Blanton E. \inst{c}
}
\institute{
\saoname
\and Sternberg Astronomical Institute, Universitetskii pr. 13, Moscow,
119899,Russia
\and Columbia University, New York, USA
}

\maketitle
\begin{abstract}
\small
Observations of the nebula associated with the WO star in the
galaxy IC 1613 are presented. The observations were carried out with a scanning
Interferometer Fabry-Perot  in \Ha at the 6m Special Astrophysical
Observatory telescope; narrow-band \Ha and [OIII] images were obtained
with the 4-m KPNO telescope. The monochromatic \Ha image clearly reveals a
giant bipolar shell structure outside the bright nebula S3. The sizes of the
southeastern and northwestern shells are $112\times 77$ pc and
$(186-192)\times(214-224)$ pc, respectively. We have studied the object's
kinematics for the first time and
found evidence for expansion of both shells. The expansion velocities of the
southeastern and northwestern shells exceed $50$ and $70\km$, respectively. We
revealed a filamentary structure of the shells and several compact features in
the S3 core. A scenario is proposed for the formation of the giant bipolar
structure by the stellar wind from the central WO star located at the boundary
of a dense ``supercavity'' in the galaxy's H I distribution.
\end{abstract}

\small
\section{Introduction}

Among more than five hundred WR stars in the Local-Group galaxies, only six
objects belong to a rare group of the ``oxygen'' stars Wolf-Rayet  (WO).
According to Barlow \& Hummer (1982), they can be considered as a separate
sequence in the WN-WC-WO chain, which represents a very short final
evolutionary stage of massive stars close to a bare CO core. The WO stars are
characterizes by a ``superwind'' (with a velocity up to $5000-6000\km$ (see
Barlow \& Hummer, 1982; Torres et al.,1986; Dopita et al.,1990;Polcaro et
al., 1992) preceded by the wind at the MS and WR stages. The WO effective
temperature reaches $10^5$ K (Maeder \& Meynet, 1989;Dopita et al.,1990;
Melnik \&Malayeri ;Polcaro et al.,1991).

Of the six WO stars in the Local-Group galaxies, three were identified in our
Galaxy, one in the LMC, one in the SMC, and one in the irregular dwarf galaxy IC
1613; only two of them are associated with bright nebulae, including WO in IC
1613.

Our interest in the latter object also stems from the fact that WO is the only
WR star identified in IC 1613, although the number of WO stars cannot
exceed $\approx1\%$ of the WR population because of their short lifetime.

D'Odorico \& Rosa (1982) and Davidson \& Kinman (1982) identified the WO star in IC
1613 by broad lines typical of this class of stars in the spectrum of the core
of the bright emission nebula S3 (Sandage, 1971). The star's coordinates
 $\mbox{RA}(1950)= 1^{h} 02^{m} 27.\!\!^{s}3$, $\mbox{DEC}(1950)=+01^\circ 48'17''$,
 are known with a $1''$ accuracy
(Armandroff \& Massey , 1985).

The central part of S3 is characterized by bright HeII 4686 emission
(Smith, 1975; D'Odorico \& Rosa, 1982; Garnett et.al, 1991).
 A detailed spectral analysis and a discussion of the chemical composition
of the bright nebula are presented in (Garnet et al., 1991, Kinsburgh \&
Barlow, 1995). Narrow-band \Ha images
revealed regions of weaker emission predominantly southeast of the
previously known  bright, elongated nebula S3 (Hodge et al., 1990; Hunter et
al., 1993).

Thermal radio emission was detected from S3; the sizes of the radio source at
half maximum intensity are $19\times 14''$, and the radio image faithfully reproduces
the shape of the bright optical nebula (Goss \& Lozinskaya, 1995).

In this paper, we present our \Ha observations of the nebula with a
Fabry-Perot interferometer on the 6m Special Astrophysical Observatory (SAO)
telescope, as well as narrow-band \Ha and [OIII] images of the region
obtained with the 4-m KPNO telescope.

For the first time, we have clearly revealed a weak outer bipolar shell
structure far outside the bright nebula S3 and studied the kinematics of the
object.

In Sect. 2, we provide basic parameters of the instruments and describe the
observing and reduction techniques; results of the interferometric
observations with the
6m telescope and the observations with the 4-m KPNO telescope are presented in
Sect. 3. The origin of the unique bipolar shell structure of the nebula around
the WO star is discussed in Sect. 4.

\section{Observations and data reduction}

\subsection{Interferometric Observations}

The interferometric \Ha observations were carried out with a scanning
Fabry-Perot interferometer  with focal reducer at the prime focus of the
6m SAO telescope SAO  (the equivalent focus is F/2.4). The
interferometer is described in Dodonov et al. (1995).
The pre-monochromatization was performed by
using an interference filter with the $FWHM= 15 \mbox{\AA}$ centered on
\Ha and tilted at an angle of $6.\!\!^\circ5$.

We used a Fabry-Perot etalon operating in the 235th order at the \Ha
wavelength; the separation between adjacent orders $\Delta\lambda = 27.9\mbox{\AA}$
corresponded to the $1274 \km$ range free from order overlapping. The
detector was a TK1024 $1024 \times 1024$-pixel CCD array. The observations were carried
out with a $2 \times 2$-pixel instrumental binning to reduce the readout time. We
obtained $522 \times 522$-pixel images in each spectral channel, with the pixel size
corresponding to 0\farcs68.

The size of the spectral channel was $\delta\lambda = 1.16 \mbox{\AA}$ or $53 \km$
in \Ha. The actual instrumental profile of the interferometer was
determined from night-sky lines $6568.8, 6577.2/6577.4, 6562.8 \mbox{\AA}$, and
\Ha); the profile width was about $3 \mbox{\AA}$ or $140 \km$.

We performed the observations on August 31, 1998,  with a $2.5-3''$ seeing at the
zenith distance $z = 51-54^\circ$. A total of 24 interferograms were obtained for
various interferometer plate spacings. The time of a single exposure was 240 s.

The He I ($6678.15 \mbox{\AA}$) line of a calibration lamp was used for the
phase shift measurement. After the observations, we made several accumulations
of the images
\begin{figure*}
\psfig{figure=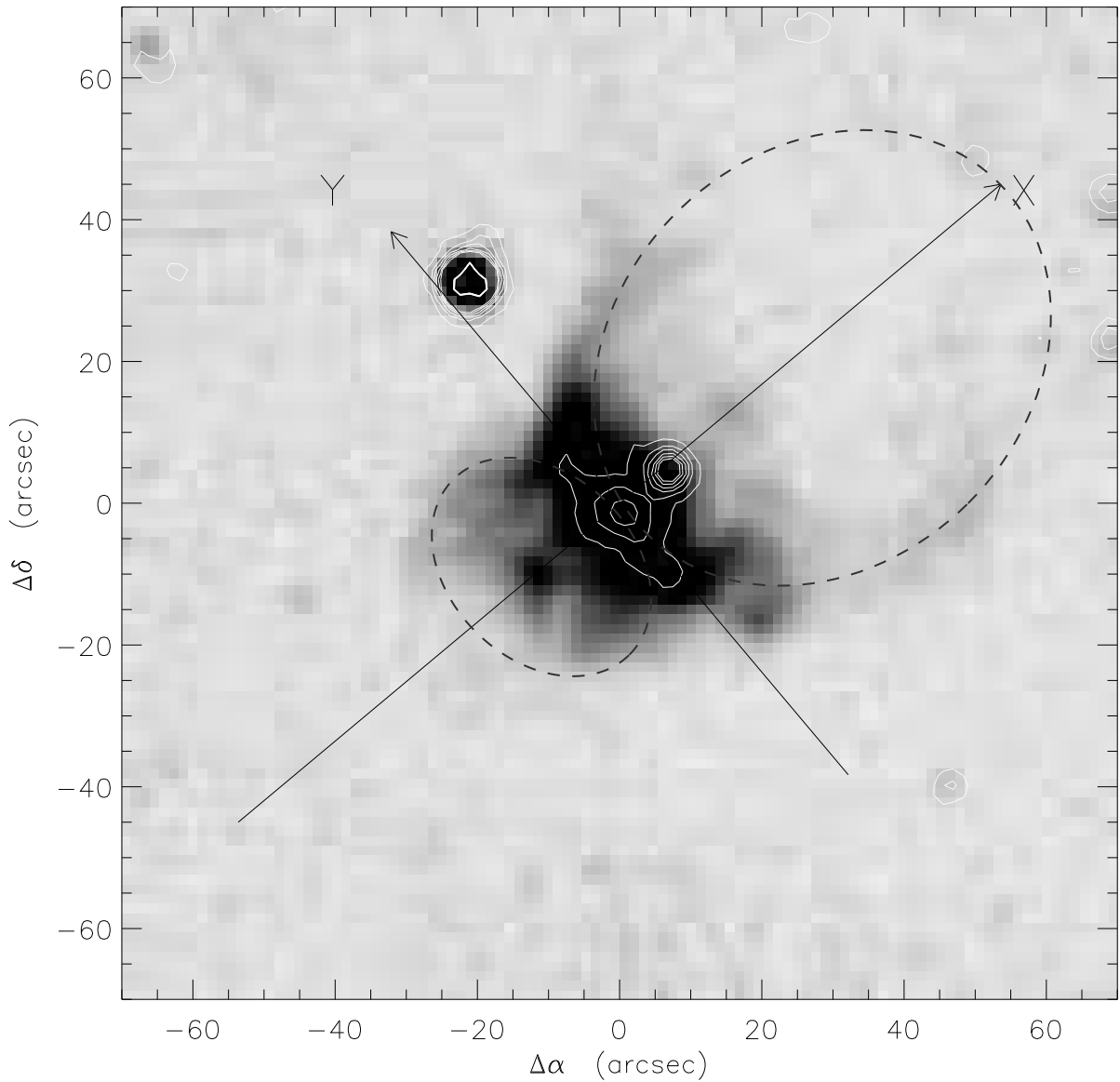,width=16 cm}
\caption{ The monochromatic \Ha image of the region obtained with
Interferometer Fabry-Perot by integrating
the emission over all spectral channels of the line. The image of the same
region in continuum near \Ha   line is indicated by isophotes. The
dashed line represents the giant bipolar shell.
}
\end{figure*}
of the caps of the 6m primary mirror uniformly illuminated by a lamp of
continuous light. This allowed us to make a correction for nonuniform detector
sensitivity (flat-fielding) and simultaneously to adjust the filter passband.

The observational data were represented as $522 \times 522 \times 24$-pixel
``data cubes'';
i.e., a 24-channel spectrum corresponded to each image element.

The data reduction (correction for the phase shift, subtraction of the night-sky
spectrum, construction of the velocity field and the  emission line and
continuum  images) was carried out by using the ADHOC software package developed
by Boulesteix (1993).

Using star images from the Digital Palomar Sky Survey Atlas (DSS), we  rotated
the original images to coincide with the correct orientation
$\alpha-\delta$ and then averaged a $2 \times 2$ element, so that the resulting
pixel size was 1\farcs34.

To increase the signal-to-noise ratio, we performed an optimal filtration of the
cube of observational data: smoothing by Gaussian with 1.5 channels FWHM in
spectral direction and smoothing in spatial direction with FWHM about of the
seeing value.
After the smoothing, the resulting spatial resolution was $4''$.

Although the actual spectral resolution for the etalon in the 235th order of
interference was $130-150\km$ in \Ha, the accuracy of measuring the
velocity from the line centroid (or from the Gaussian fit) in the case of a
symmetrical profile is considerably higher and is determined by the accuracy of
phase calibration. The actual accuracy of a single measurement was checked by
using a comparison spectrum and was $8-10 \km$.

The tie-in to the absolute velocity  was checked by using night-sky lines. The
possible systematic shift did not exceed $15-20\km$.

\subsection{Narrow-Band Images}

The narrow-band \Ha and [OIII] images of the nebula were obtained with
the 4-m Kitt Peak National Observatory (KPNO) telescope. Interference filters
centered on the \Ha ($\lambda= 6562 \mbox{\AA}$, $FWHM = 29 \mbox{\AA}$)
and [OIII] ($\lambda = 5009 \mbox{\AA}$, $FWHM = 45 \mbox{\AA}$) lines were used.
The detector was a $2048 \times 2048$-pixel CCD array with the pixel size corresponding
to 0\farcs47.

The observations were carried out on December 29 and 30, 1995; the angular
resolution determined from star images was 1\farcs2 in \Ha and 1\farcs4 in [OIII].

\section{Results of the observations}

\subsection{Monochromatic \Ha Image}

Figure 1 shows the monochromatic \Ha image of the region obtained by
integrating the emission over all spectral channels of the line. For comparison,
the image of the same region in continuum in the immediate vicinity of \Ha
 line is indicated by isophotes; here,  foreground stars in the field of view
are clearly
seen.

The absolute brightness were determined from comparison with the brightness
from  Hodge et al. (1990) for several nebular regions. Sources 37b, c, e,
and g in the list of these authors were used for the flux calibration;
i.e., the brightest and faintest regions were excluded.

Constructing the monochromatic image allowed us to clearly distinguish regions
of weak \Ha emission from the background far outside the bright nebular
core. As a result, we have been able for the first time to clearly show that
\begin{figure*}
\psfig{figure=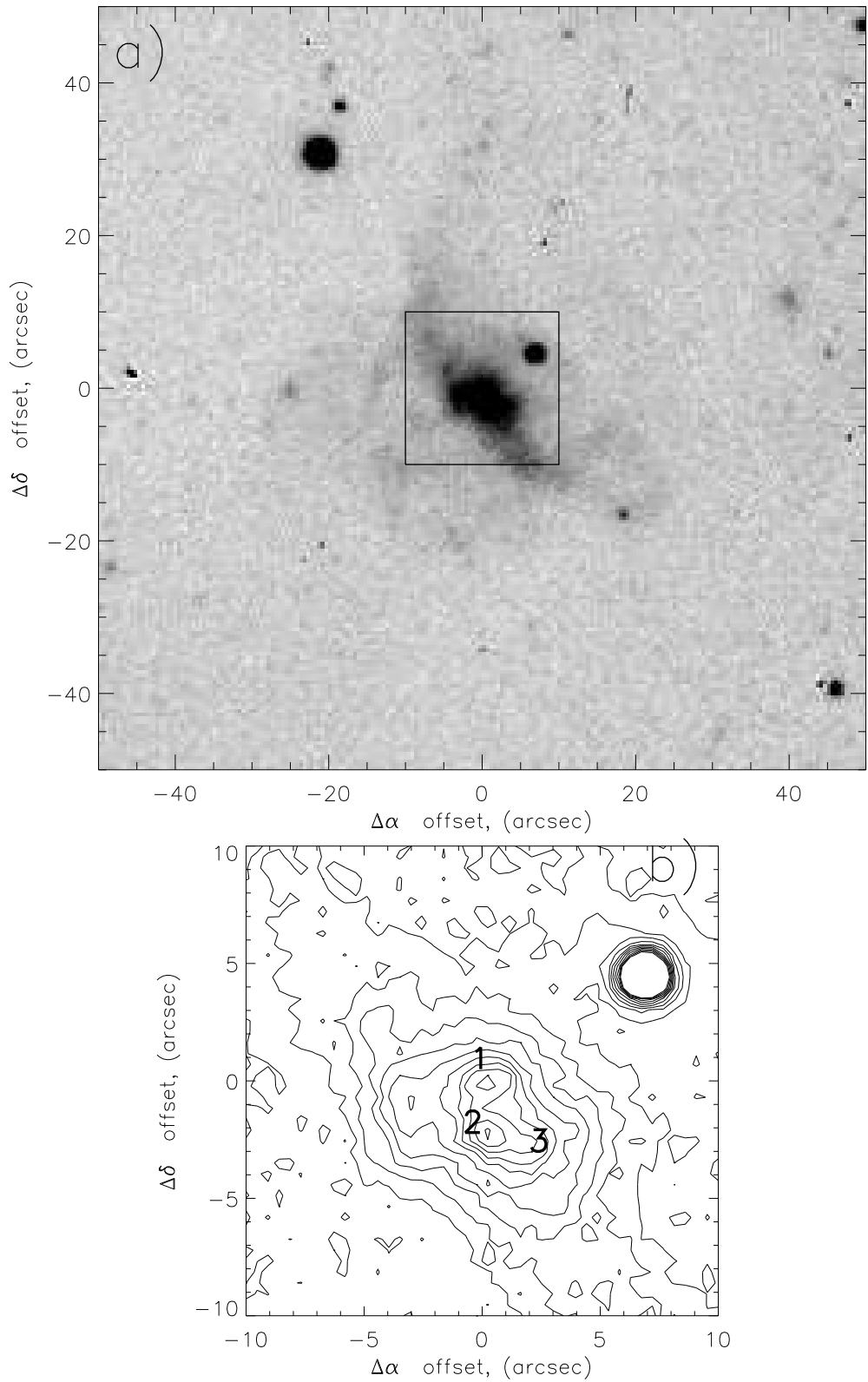,width=14 cm}
\caption{
The   narrow-band \Ha (a, b) and [OIII] (c, d) images of the
nebula obtained with the 4-m KPNO telescope. The numbers denote three compact
sources; source N1 is most likely the WO star. The square encloses the region
whose isophotes are shown below.
}
\end{figure*}
these outer weak regions form a giant bipolar shell structure around the WO star
and the bright S3 core. The two shells forming this bipolar structure are
indicated by the dashed line in Fig. 1.

According to our measurements, the sizes of the three  components  (the bright
core and the two outer shells) are the following: angular sizes of
$26\times 20''$ or linear sizes of $83\times66$ pc for the bright nebula
S3; $35\times 24''$ or $112\times77$ pc
for the outer southeastern shell; and $(58-60)''\times (67-70)''$ or
$(186-192)\times(214-224)$ pc for the outer northwestern shell.

We give the sizes along the major and minor axes of each structure for a
distance of 660 kpc, as inferred by Saha et al. (1992)

\subsection{Narrow-Band \Ha and [OIII] Images}

Figure 2 shows the \Ha (Figs. 2a, 2b) and \mbox{[OIII]} (Figs 2c, 2d) images of
the nebula obtained with the 4-m KPNO telescope. The southeastern outer shell
and part of the northwestern shell at the very ``base'' near the bright nebular
are seen in these figures. Both shells exhibit a distinct filamentary structure;
in general, the filaments in the two lines are morphologically identical.

Three compact features unresolvable in the  6m telescope
monochromatic image because of the
poor seeing on the observing night are clearly seen in the region of the bright
S3 core. In all probability, the brightest of them (N1 in Fig. 2) is the WO
\begin{figure*}
\psfig{figure=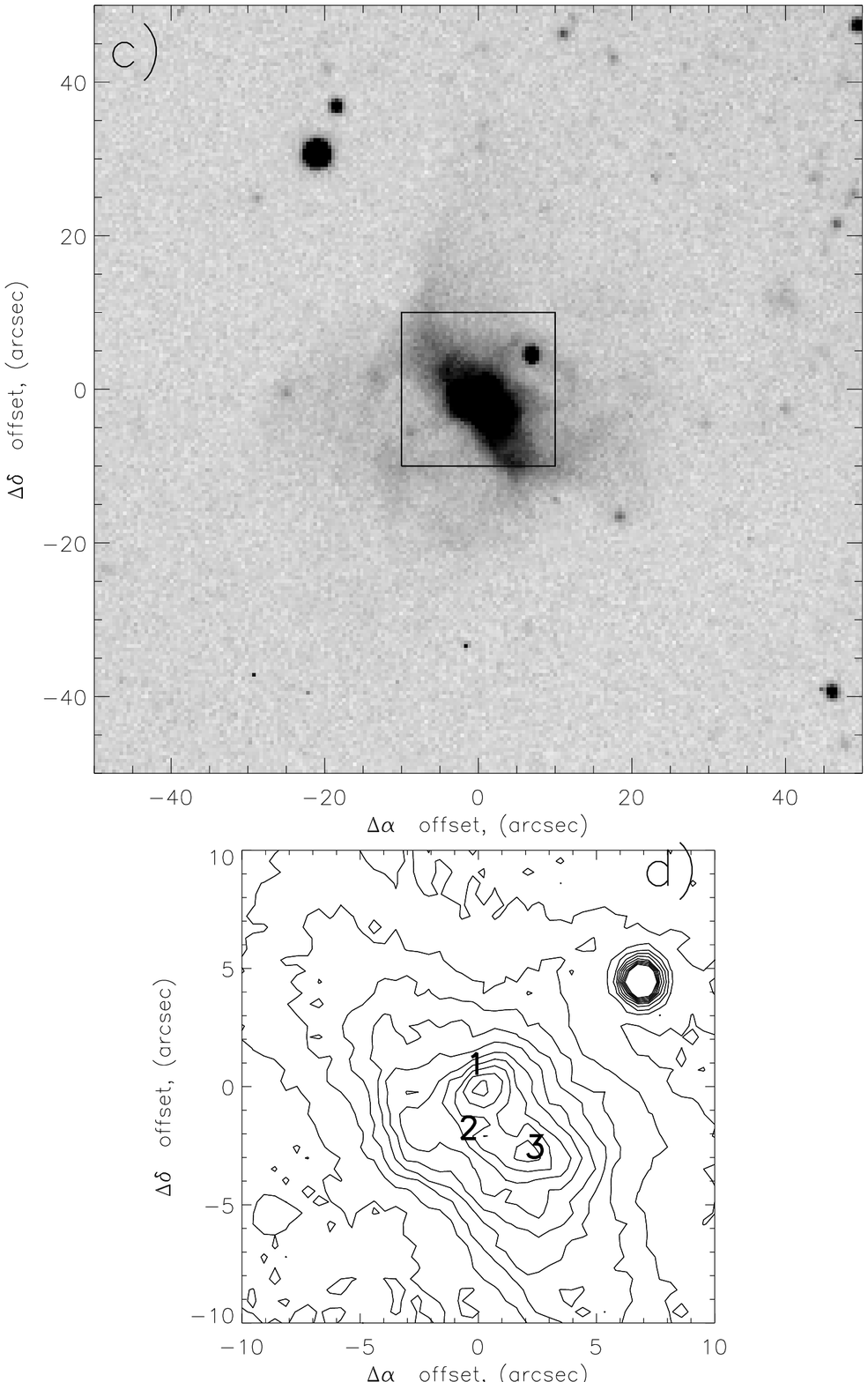,width=14 cm}
Figure 2. continued
\end{figure*}
star. The remaining features are most likely compact knots of gas, because, as
follows from a comparison of Figs. 2b and 2d, their relative brightnesses in the
two lines are different (N2 is brighter than N3 in \Ha and weaker than the
latter in [OIII]). To elucidate the nature of these compact sources requires
spectroscopic observations at a sufficiently good seeing.

\subsection{Kinematics}

Figure 3 shows the measured velocities at the line peak and the line FWHMs as
 obtained from our Fabry-Perot interferometric observations. In the data
reduction, the line was fitted by a Gaussian; the line width was corrected for
the instrumental profile width by assuming that the latter was also Gaussian. In
order to find evidence for expansion of the outer bipolar shell, we constructed
the distributions of line velocity and FWHM along two cuts designated,
\begin{figure*}
\mbox{
\psfig{figure=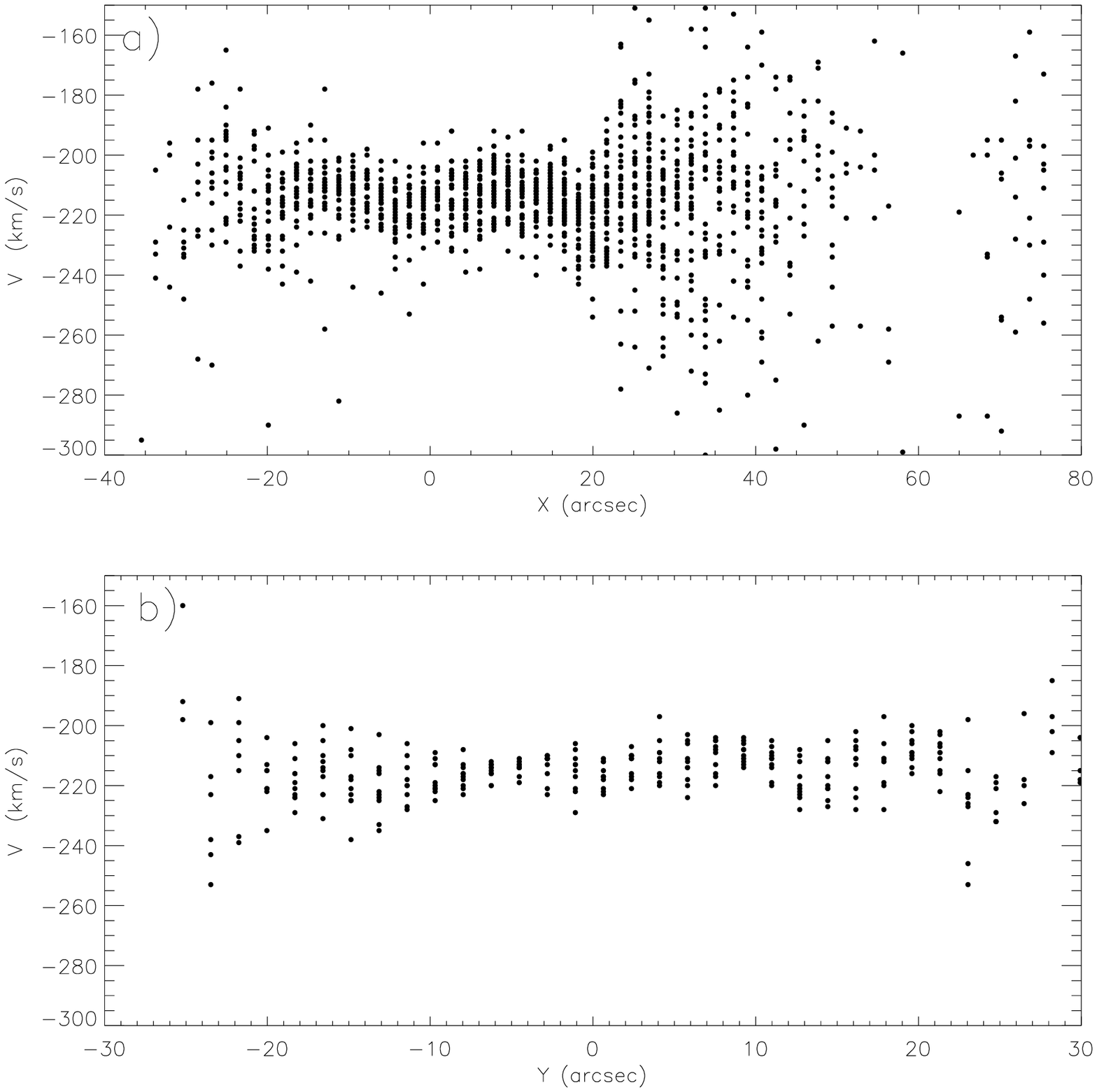,width=8.2 cm}
\psfig{figure=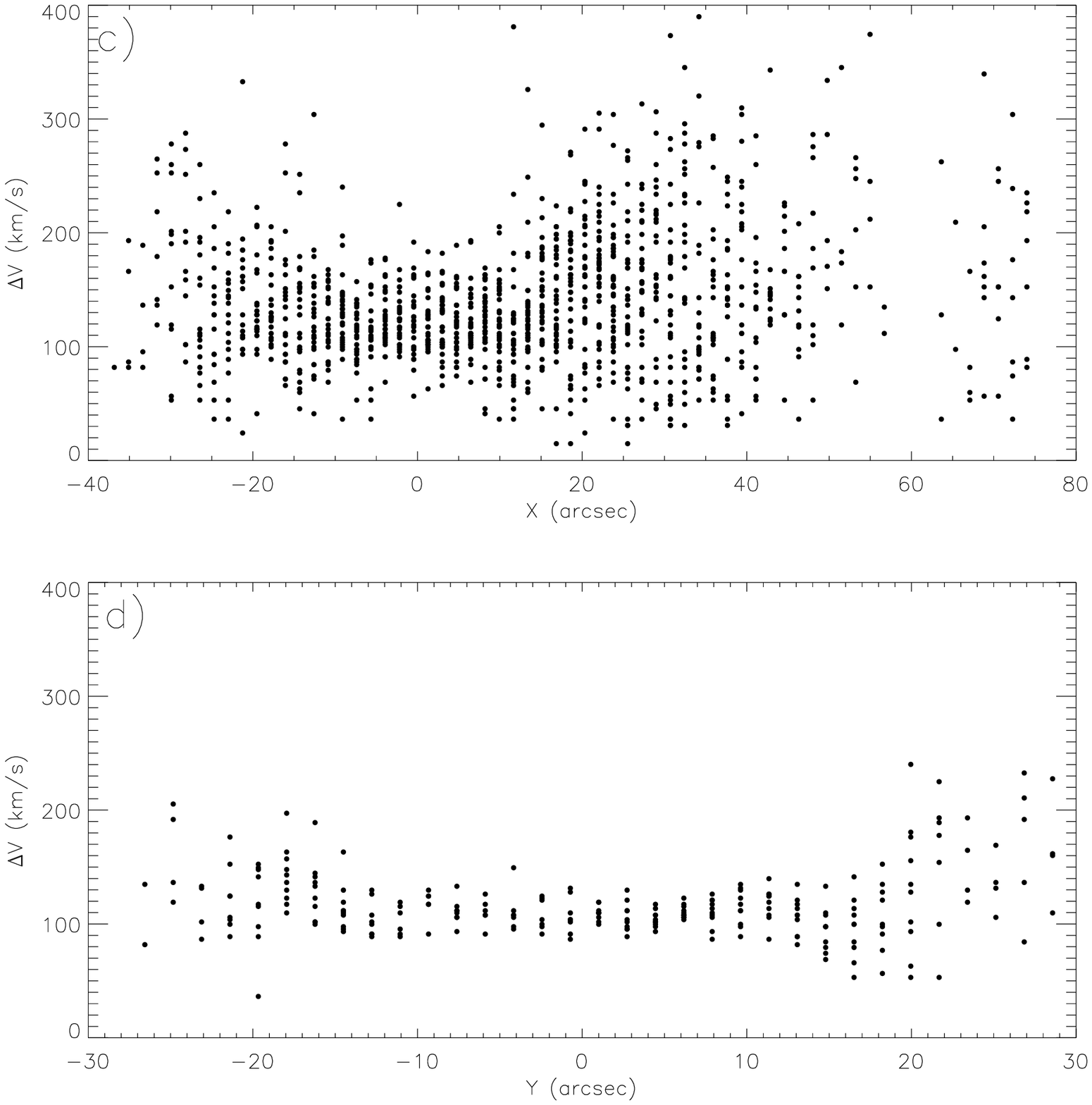,width=8.2 cm}
}
\caption{The measured velocity at line peak and the line FWHMs  from
our Fabry-Perot interferometric observations: (a) $V(Hel)$ in a $60''$ wide band
along the ``X''-axis (see Fig. 1); (b) $V(Hel)$ in a $15''$ wide band along the
``Y''-axis; (c) and (d) the line FWHMs in the same bands along the ``X''
and ``Y'' axes, respectively.
}
\end{figure*}
respectively, ``X''  and ``Y'' in Fig. 1. The band along the ``X'' axis is
$60''$ wide, so the entire bipolar structure falls within it; the ($15''$
wide) band along the ``Y''
axis covers predominantly the bright central S3 core and, in part, the weaker
emission at the bases of the two outer shells.

A comparison of the line velocities and FWHMs along the two cuts leads us to the
following conclusions:

\begin{enumerate}
\item The scatter of velocities of the line peak is systematically larger in both
outer shells than that in the central bright S3 core. Indeed, the velocities in
the central region $(-10''\div+10'')$ along the ``X'' axis in Fig. 1 vary in the
range from $-190$ to $-255\km$; the velocities in the southeastern shell
vary in the range from $-165$ to $-295 \km$; and the range of velocities in
the extended northwestern shell is largest: from $-150$ to $-300 \km$.

As we see from the cut along the ``Y'' axis in Fig. 3b, the scatter of velocities
in the bright S3 core is even smaller: velocities from $-200$ to $-230 \km$
are observed in a band within $\pm13''$ of the center (in accordance with the core
size; see Fig. 1). At large distances from the center, the scatter increases
because of the contribution of emission from the bases of both outer shells.

\item The line width in both shells also exceeds appreciably the line width in the
central core. As we see from Fig. 3c, the line width in the same central region
$(\pm10'')$ is no larger than $240 \km$; in the southeastern shell, it reaches
$330\km$; and the broadest line up to $390\km$ is observed in the
northwestern shell.
\end{enumerate}

The two observed patterns of behavior suggest the presence of systematic
high-velocity motions in the outer bipolar shell structure. Based on these data,
we may take an expansion velocity $\ge 50 \km$ for the southeastern shell and
$\ge> 70 \km$ for the northwestern shell as a rough estimate. The two values
are a lower limit; the underestimate can be more significant for the
northwestern shell, because here we observe mainly the emission from the
periphery, where the effect of geometrical projection reduces the observed
radial velocities.

Evidently, observations with a higher spectral and angular resolution are
desirable to accurately estimate the expansion velocity. It should be borne in
mind, however, that such an ``erratic'' velocity distribution -- from slow bright
knots to the most rapid weak features -- is observed in most expanding
supernova remnants and nebulae around the sources of strong stellar wind, which
were studied with a high angular and spectral resolution (Lozinskaya,1986
and 1992). Indeed, the
nebulae are irregular in shape and clumpy in structure; accordingly, the effect
of geometrical projection changes erratically the observed radial velocity.
During the passage of a shock wave, the gas of dense bright knots accelerates
weakly, while the gas of lower density accelerates more strongly. For these
reasons, the entire set of radial velocities, ranging from the gas at rest to
the shock velocity, is commonly observed in real expanding shells. Higher
resolution observations of the nebula around the WO star in IC 1613 could
therefore reveal the same erratic distribution of line velocities and FWHMs.

According to our measurements, the mean velocity of the bright part of the
nebula, where the expansion effect is marginal, is $V(Hel) = -216\km$.
Given a possible systematic tie-in error of the absolute velocity measurements
within $15-20 \km$ (see Sect. 2), this value is in agreement with the
measurements by Tomita et al.(1993): $V(Hel) = -235 \div -230 \km$, as
inferred from one spectrogram of S3, and with the measurements by Lake \&
Skillman (1989) in the 21-cm radio line: the mean velocity $V(Hel) = -232\km$
in the near galactic region.

\section{Discussion}

The existence of an outer bipolar structure outside the bright core of the
nebula S3 in IC 1613 was first suspected by one of us (T.A.L.) back in 1988 on
the basis of trial narrow-band [OIII] images obtained at her request by M.A.
Dopita with the 2.6m MSSS ANU telescope, which have not been published. This
suspicion was completely confirmed in this study.

Lozinskaya (1997) suggested the   break out  (burst) of a strong stellar wind in
two directions out of the dense gas layer forming the ``main body'' of the nebula
as a possible explanation for the bipolar structure of the outer shells. Not
only the morphology of the nebula but also its orientation relative to the
\begin{figure*}
\psfig{figure=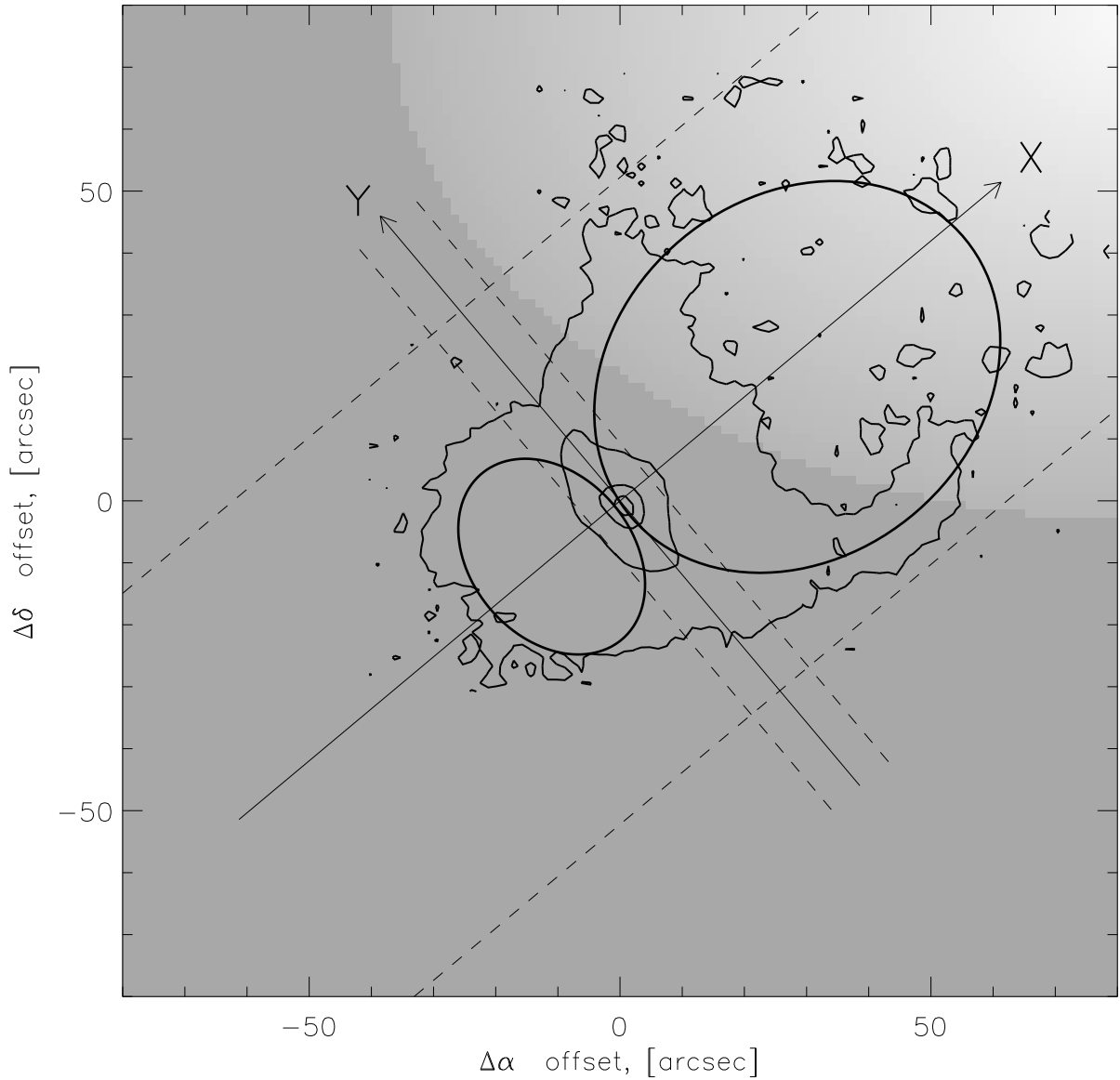,width=16 cm}
\caption{ A scheme for the bipolar shell structure associated with the WO
star and its location relative to the HI supercavity. The bipolar shell is
represented by two ellipses. The dashed lines indicate the boundaries of
the cuts along the ``X'' and ``Y'' axes (see Fig. 3). The \Ha isophotes as
constructed from the 6m
telescope observations correspond to 0.04, 0.5, 1.5 and
$2.0\times10^{-15}\mbox{erg}\,\mbox{cm}^{-2}\,\mbox{s}^{-1}/\Box''$
The shades of gray indicate the distribution of HI column
density (the fit to the data from Lake \&  Skillman, 1989).
The transition from white to gray
corresponds to column densities from
$0.4$, to $2\times10^20\,\mbox{cm}^{-2}$
}
\end{figure*}
extended region of H I deficiency, which is clearly seen in the HI
distribution from the 21-cm line emission obtained by Lake \& Skillman
(1989), argue for this model. The center of this ``supercavity'' in the H I distribution
has the following coordinates: $\mbox{RA} = 1^h2^m20^s$ and $\mbox{DEC}=
1^\circ 51'$; its size is $2-3'$ ($400-600$ pc). The formation of such a
supercavity may be associated with the preceding activity of stars in the
region. If the H I deficiency is assumed to be caused by the sweeping up of
gas, then the orientation of the nebula's main body along the boundary of
the hypothetical swept-up H I ``supershell'', the southeastern shell in the
region of a denser medium, and the extended weak northwestern shell in the
medium of lower density inside the supercavity can be explained naturally.

The bipolar shell structure associated with the WO star in IC 1613 and its
location relative to the H I supercavity are schematically shown in Fig. 4.

The analogy with an extensively studied galactic object also argues for the
break out of a strong stellar wind into the region of reduced gas
density: the bright shell swept up by the wind in a dense cloud and the
``blister'' produced by the wind burst through the cloud into a tenuous medium are
also observed in the nebula G2.4+1.4 associated with the star of the same type,
WR 102 (Dopita \& Lozinskaya, 1990; Dopita et al.,1990).

An analysis of the parameters of the two shells associated with the WO star in
IC 1613 yields the rough estimates that are in best agreement with the standard
model of wind effect on the interstellar medium. The size of both shells
($R_1 \simeq56$ pc and $R_2 \simeq 110$ pc) is actually the only parameter that was reliably
determined from observations. However, the fact that the ages of the two shells
must be the same in the proposed model constraints significantly the range of
possible values for the parameters.

The wind  mechanical luminosity for the WO star in IC 1613 is
$L_{w} \simeq 10^{38}\erg$ (Kinsburgh \& Barlow, 1995).
 This estimate is rather crude, because only the wind velocity
 $(V_w =2850 \km)$ was determined from observations, while the mass-loss rate
 ($2.9\times10^{-5}$ M$_\odot/\mbox{yr})$ was estimated indirectly from the
 relation between the HeII-line luminosity and the outflow rate for two WO
 stars.

The typical density of the interstellar medium in the vicinity of the star can
be estimated with the same uncertainty.

First, based on the $\mbox{H}_\beta$ emission from the brightest central part of
S3 $6''$ in size, Davidson \& Kinman (1982) found the core density to be
$N_{e}=8.5~\mbox{cm}^{-3}$.
Second, using the total \Ha emission from the entire bright nebula,
Kennicutt (1984) obtained a mean $N_{e}\simeq1~\mbox{cm}^{-3}$. Third, Goss
\& Lozinskaya (1995) estimated the mean density in the bright nebula S3
from the flux density of  thermal radio emission to be $N_{e}=3.5~\mbox{cm}^{-3}$.

Given the dense knots that we detected in the central core, all the three
estimates give a lower limit on the density in the bright nebula. These
estimates refer to the densest region of the central nebula, where, judging by
the radial-velocity field presented above, the gas was not accelerated by the
stellar wind because of the high density.

The mean ambient gas density in a large region around the WO star can be
obtained from the 21-cm observations by Lake \& Skillman (1989). Their map of
the H I column-density distribution gives $\mbox{N(HI)}\simeq (2-4)\times
10^{20}~\mbox{cm}^{-2}$ for
the region of interest. For a thickness of the galactic gas disk of
$500-800$ pc, this value corresponds to a mean density of $\simeq
0.1~\mbox{см}^{-3}$. We estimated the
gas-disk thickness from the following considerations. According to the optical
B surface-brightness profile given in the above paper, the stellar surface
density at a distance $R\approx6'$ from the center is
$\sigma_*\approx2.5-3~ M_\odot/\mbox{pc}^2$,
while the total surface density of stars and gas is
$\sigma_{T}\approx3.5-4~ M_\odot/\mbox{pc}^2$.
 The thickness of the gas disk can be roughly assumed to be
$h_0\sim1/\sqrt\sigma_{T}$ (see, e.g. Zasov,1993 ). Since
$\sigma_{T}\approx50-70~M_\odot/\mbox{pc}^2$ at a
gas-disk thickness of $\approx200$ pc in the solar neighborhood, we obtain
$h_0\approx 500-800$ pc in the WO region for IC 1613.

All these estimates are rather crude, but they show that the initial density in
the vicinity of the WO star is unlikely to fall outside the range from
$0.01$ to $5~\mbox{cm}^{-3}$.

The expected ages (in units of $10^{6}$ years) and expansion velocities for the two
outer shells in the standard model   of a wind-blown bubble (Castor et
al.,1975; Weaver et al.,1977) are given
in the table  for various initial densities $n_0$ at $L_w = 10^{38}\erg$
and $R_1\simeq56$ pc and $R_2\simeq110$  pc derived above.

\begin{table}
\caption{The expected ages and expansion velocities of the two outer shells in a medium
of various densities}
\begin{tabular}{|c|c|c|c|c|c|}
\hline
   $n_o, \mbox{cm}^{-3}$  &    5    &   1      &   0.1    &  0.01   & 0.001  \\
   \hline
   $t_{6}(R_{1})$ &   1.2   &  0.7     &   0.3    &  0.15   & 0.07   \\
   $v_{1},\km$  &   28    &  46      &  100     &  218    & 460  \\
   \hline
   $t_{6}(R_{2})$ &   3.6   &  2.1     &   1.0    &  0.45   & 0.2  \\
   $v_{2},\km$  &   18    &  30      &   63     &  140    & 300  \\
\hline
\end{tabular}
\end{table}

As we see from the table, similar age estimates for the two shells, depending on
the radius, provided that the expected expansion velocity is close to the
observed values $v_1\ge 50\km$ and $v_2\ge 70\km$, are obtained at an
initial gas density $n_o\simeq 1-0.1 \mbox{cm}^{-3}$ for the brighter
southeastern shell and $n_o\simeq 0.1-0.01\mbox{cm}^{-3}$ for the weaker
extended northwestern shell. These ranges of ambient gas densities seem
quite reasonable. The ratio of ambient gas densities in the two shells
swept up by the wind is determined from the ratio of their sizes more
accurately than the density itself:
$n_{o,1}/n_{o,2}=(R_{2}/R_{1})^5\simeq20$.

The inferred age of the bipolar shell structure, $t_1\simeq(0.3-1)\times10^6$ years, is
close to the lifetime of a massive WR star and exceeds appreciably the duration
of the final WO stage. Note that the age was estimated for the WO wind
intensity. The wind intensity at the preceding WR stage was most likely lower,
which increases still further the age estimate for the shell. Thus, we conclude
that the unique bipolar structure was most likely produced by the wind from
the central star at the stage preceding WO. The WO superwind, which switches on
inside the previously formed structure, is an additional source of energy. Since
the shock wave triggered by the strong WO wind propagates in a medium of very
low density inside the existing bipolar shells, the ``secondary'' shells swept up
by them rapidly reach the denser regions at the boundary.
The additional compression of
gas by the shock wave triggered by the superwind gives rise to a filamentary
structure, which is clearly seen in Figs. 3a and 3b in the southeastern shell
and at the base of the northwestern shell.

\section{Conclusion}

Our \Ha observations of the nebula associated with the WO star in the
galaxy IC 1613 using a scanning Fabry-Perot interferometer on the 6m SAO
telescope  clearly revealed a giant bipolar shell structure outside the
bright nebula S3. The sizes of the shell southeast of the bright nebula are
$112\times77$ pc; the northwestern shell is twice as large:
$(186-192)\times(214-224)$ pc.
The object's kinematics has been studied for the first time. We found evidence
for expansion of both shells; the possible expansion velocities of the
southeastern and northwestern shells are no less than $50$ and $70 \km$,
respectively. Using the 4-m KPNO telescope (narrow-band \Ha and [OIII]
images), we detected a filamentary structure of both shells and several compact
features in the S3 core.

We propose a self-consistent model for the formation of the giant bipolar
structure by the wind from the central WO star located at the boundary of a
dense supercavity in the galactic H I distribution.

\acknowledgements{
\footnotesize
This study was supported by the Russian Foundation for Basic Research (project
no. 98-02-16032) and the Program ``Astronomy'' (project no. 1.3.1.2). We wish to
thank M. Dopita, who provided unpublished [OIII] images of the region, J.
Boulesteix for the opportunity to use interference filters, and the 6m
telescope committee for allocating observing time.}

\end{document}